\documentclass[12pt]{revtex4}

\usepackage{amssymb}

\newcommand{\tL}{\tilde{L}}
\newcommand{\La}{L_{\alpha}}
\newcommand{\tE}{\tilde{E}}
\newcommand{\Ea}{E^{\alpha}}
\newcommand{\tpsi}{\tilde{\psi}}
\newcommand{\psia}{\psi^{\alpha}}

\newcommand{\tl}{\tilde{\lambda}}
\newcommand{\la}{\lambda_{\alpha}}

\begin{document}

\title{Multi-point quasi-rational approximants for the energy eigenvalues of potentials of the
  form {\boldmath $V(x)= Ax^a + Bx^b$}}

\author{P. Mart\'in}
\author{A. \surname{De Freitas}}
\author{E. Castro}
\affiliation{Departamento de F\'isica, Universidad Sim\'on Bol\'ivar,
Apdo. 89000, Caracas, 1080 A, Venezuela}
\author{J. L. Paz}
\affiliation{Departamento de Qu\'imica, Universidad Sim\'on Bol\'ivar, 
Apdo. 89000, Caracas, 1080 A, Venezuela}

\begin{abstract}
Analytic approximants for the eigenvalues of the one-dimensional
Schr\"odinger equation with potentials of the form $V(x)= Ax^a + Bx^b$ are found
using a multi-point quasi-rational approximation technique. This technique is
based on the use of the power series and asymptotic expansion of the
eigenvalues in $\lambda=A^{-\frac{b+2}{a+2}}B$, as well as the expansion at intermediate
points. These expansions are found through a system of differential
equations. The approximants found are valid and accurate for any value of $\lambda$.
As examples, the technique is applied to the quartic and sextic 
anharmonic oscillators.
\end{abstract}



\maketitle

\section{Introduction}

There are many techniques that have been developed with the purpose of solving
the Schr\"odinger equation since it was first introduced more than eighty
years ago. One might think that after so many years this should be a closed subject,
yet this area of research is still being pursued nowadays by a number of 
physicists in the world. There are still many potentials of interest for which no 
exact solution is known and,
moreover, new potentials that seem to model the behavior of
physical systems, such as a set of different molecules interacting with each
other or with an external field, are still being proposed from time to
time, which prompts physicist to study them in more detail. A
recent example of this can be found in \cite{LeRoy1}. Since one cannot find exact
solutions for many of the most interesting potentials, one is left
with one of two options: (1) use a numerical method or (2) try to find an
approximate analytic solution. The last option is particularly appealing,
since in many situations it is possible to use precise approximate solutions in
the same way as the exact ones. 
Approximate analytic solutions can be
obtained by different methods for both, the energy eigenvalues and eigenfunctions
of the Schr\"odinger equation, although more methods can be found in the
literature for the eigenvalues than for the eigenfunctions. In either
case, there are hundreds of publications devoted to this subject, and it would
be impossible to make justice by citing them all. 

An analytic approximant should be a function of the parameters of the potential
that comes very close to the values of the exact solutions (found numerically) when
evaluated at any particular point in the parameter space. The usefulness of a
particular method used to obtain these approximants will depend on the
precision of the approximations as well as the simplicity of the analytic expressions.

In this paper, a new method is proposed for finding analytic approximants to the energy
eigenvalues of the one-dimensional Schr\"odinger equation with potentials of
the form $V(x)=Ax^a+Bx^b$ \cite{R1,R2,R3,R4,R5,R6}. This kind of potentials are perhaps not the most
interesting ones from a phenomenological point of view, but they are
certainly of theoretical interest, since they include such potentials as the
quartic and sextic anharmonic oscillators, and in any case, they are a good choice
to test new methods before applying them to more interesting problems.

The method we are proposing here is based on the multi-point quasi-rational
approximation technique, which has been successfully applied before to
similar kinds of problems involving differential equations \cite{CM1,CM2,CM3,M1,CM4,CM5}.
This technique consists in using the expansions of the
function to be approximated around different values of the parameters in the
differential equation where this function appears, in order to write an
approximant in terms of rational functions in these parameters combined with
auxiliary ones. The approximant will then have almost the same expansions
around the different chosen values of the parameters. The auxiliary functions
are usually needed in order to match the behavior of the quantity to be
approximated when the parameters go to infinity, which normally cannot be done
solely using rational functions as in a Pade's approximation.

In our case, the differential equation we are interested in is, of course, the
Schr\"odinger equation
\begin{equation}
\left(-\frac{d^2}{dx^2} +Ax^a+Bx^b \right) \psi = E\psi \,\, .
\end{equation}
In this case, we have too parameters, $A$ and $B$, which will be assumed to be
both positive to simplify the treatment, though the method can be used without
this restriction. It is also assumed that $a$ and $b$ are positive integers, and
$b>a \geq 2$. As it is very well
known \cite{Simon1}, one can make this equation to depend on only one parameter by
making the changes, $x = A^{-\frac{1}{a+2}}x'$ and 
$E' = A^{-\frac{2}{a+2}}E$, which leads to
\begin{equation}
\left(-\frac{d^2}{dx'^2} + x'^a+\lambda x'^b \right) \psi = E'\psi \,\, ,
\label{SE}
\end{equation}
where $\lambda=A^{-\frac{b+2}{a+2}}B$.
From now on we will drop the primes and rename $x' \rightarrow x$ and $E'
\rightarrow E$. The energy eigenvalues $E$ will depend on the parameter
$\lambda$. Our goal is to find an approximating function
for $E(\lambda)$ for each energy level, using expansions around different values of $\lambda$,
including the power series (perturbative expansion around $\lambda=0$), and the
asymptotic expansion ($\lambda \rightarrow \infty$). In sections
\ref{PowerSeries} and \ref{AsymptoticExpansion} a neat
way to find these expansions will be shown using a system of coupled differential
equations, which in the case of the power series provides an interesting
alternative to standard perturbation methods, since here the perturbed
eigenvalue can be found using only the corresponding unperturbed state,
instead of using the whole unperturbed eigenvalue spectrum as in the usual
quantum mechanics perturbation theory. In sections \ref{quartic} and \ref{sextic}, the
construction of the approximants will be shown, using the quartic and sextic
anharmonic oscillators as examples. These approximants will be found for the ground state
energy eigenvalues, as well as the first and second excited levels. The important point of our
technique is that the same approximant will be valid and accurate for any value of
$\lambda$ (including large and small values).

\section{Power Series}
\label{PowerSeries}

The expansion of the energy eigenvalues and eigenfunctions around $\lambda=0$ can be written as
\begin{eqnarray}
E &=& E_0+E_1\lambda+E_2\lambda^2 + \cdots \,\, ,   \\
\psi &=& \psi_0 +\psi_1\lambda +\psi_2\lambda^2 + \cdots \,\, .
\end{eqnarray}
One would like to find the coefficients $E_0$, $E_1$, $E_2 \ldots$.
This can be done introducing these expansions in equation (\ref{SE}), and
demanding it to be satisfied at every order in $\lambda$, which leads to the 
following system of differential equations
\begin{eqnarray}
L\psi_0 &=& E_0 \psi_0 \,\, , \label{PS_DE1} \\
L\psi_1 + x^b \psi_0 &=& E_0 \psi_1 + E_1 \psi_0 \,\, , \label{PS_DE2} \\
L\psi_2 + x^b \psi_1 &=& E_0 \psi_2 + E_1 \psi_1 + E_2 \psi_0 \,\, ,\\
\vdots \phantom{x^b \psi_1}  && \phantom{E_0 \psi_1} \vdots \nonumber \\
L\psi_n + x^b \psi_{n-1} &=& \sum_{k=0}^n E_{n-k} \psi_k \,\, , \label{PS_DEn}
\end{eqnarray}
where
\begin{equation}
L = -\frac{d^2}{dx^2} + x^a
\end{equation}
It is important to note that, since $\lambda$ is arbitrary, 
many of the properties of the eigenfunction
$\psi$ will be inherited by the functions $\psi_0$, $\psi_1 \ldots$ . 
For example, given that for bound states, the function $\psi$
should fall off quickly for large values of $x$, so should the expansion
functions $\psi_k$. Also, if the function $\psi$ has definite parity, then the
functions $\psi_k$ will all have the same parity as $\psi$.

The coefficients in the expansion of the energy can be found by
solving numerically the differential equations one by one (using, for example,
the shooting method). For
instance, one could solve equation (\ref{PS_DE1}), obtaining then a
numerical value for the
coefficient $E_0$, together with a numerical solution for $\psi_0$. 
For a bound state, one should find a solution for a range of values in $x$ where
the fall off to zero of $\psi_0$ can be seen. Then one can
make a very precise fit (a large polynomial in $x$) for $\psi_0$ within this
range, and use this fit as an input to solve equation (\ref{PS_DE2}). This
then leads to a numerical value for $E_1$, and a numerical solution for
$\psi_1$, and in principle, the procedure can be repeated until one has as many
coefficients $E_k$ as one desires. 

Of course, the applicability of this method is limited by the numerical
precision with which the differential equations are solved, and since the
method is iterative, the numerical errors from the first $n$ equations will
be propagated to the solution of equation $n+1$. For this reason, one would
expect the precision of $E_k$ to be lower for larger values of $k$.

The numerical values of the coefficients $E_k$ can also be found directly
using the solutions of the first $k-1$ differential equations,
without solving the $k$-th one. For example, if one has
already obtained $E_0$ and $\psi_0(x)$, one can multiply both sides of equation
(\ref{PS_DE2}) by $\psi_0(x)$, and integrating in $x$ it is possible to show
that
\begin{equation}
E_1 = \frac{\int_{-\infty}^{\infty} dx x^b \psi_0^2}{
      \int_{-\infty}^{\infty} dx \psi_0^2} \,\, ,
\label{intE1}
\end{equation}
which coincides with the expression obtained using standard perturbation
methods. The same procedure can be repeated for all the other equations, leading to
\begin{equation}
E_n = \frac{\int_{-\infty}^{\infty} dx \left( x^b \psi_{n-1} 
                           -\sum_{k=1}^{n-1} E_{n-k} \psi_k \right) \psi_0}{
      \int_{-\infty}^{\infty} dx \psi_0^2} \,\, .
\label{intEn}
\end{equation}
This way of finding $E_k$ is more precise, since the number of
differential equation to be solved is $k-1$. In practice, the integrals are taken within
a range of $x$ where the functions $\psi_k(x)$ have already fallen to very
small values.

On the other hand, equation (\ref{PS_DE1}) can be solved exactly for $a=2$,
since then it would be the Schr\"odinger equation for a harmonic
oscillator. It can be shown that in this case, all the other 
equations in the system can also be solved exactly. For example, for the ground state 
$E_0=1$ and $\psi_0(x) \propto \exp(-\frac{x^2}{2})$. If we take $b=4$ (quartic
anharmonic oscillator) the next function, $\psi_1(x)$, can be written as
\begin{equation} 
\psi_1(x)=(p_0+p_1x+p_2x^2+p_3x^3+p_4x^4)\exp(-\frac{x^2}{2}) \,\, .
\end{equation}
When this is introduced in equation (\ref{PS_DE2}), the function $\exp(-\frac{x^2}{2})$ disappears and
a relation between two polynomials is left. Since this relation must be
satisfied at each order in $x$, a system of equations in $E_1$ and the $p_i$'s
is obtained, whose solution is
\begin{equation}
p_1=0\, , \,\, p_2=-\frac{3}{8} \, ,\,\, p_3=0 \, ,\,\,
p_4=-\frac{1}{8} \, ,\,\, E_1=\frac{3}{4} \,\, ,
\end{equation}
and it can be seen that $p_0$ arbitrary, which means that just like for $\psi_0(0)$, 
the initial condition $\psi_1(0)=p_0$ is arbitrary (this will be the case for
all the other functions in the expansion).

The same procedure can be repeated for $\psi_2$, $\psi_3$, etc., writing
\begin{equation}
\psi_n = \left( \sum_{k=0}^{4n} p_k x^k \right) \exp(-x^2/2) \,\, .
\end{equation}
We obtain
\begin{equation}
E_0=1 \, ,\,\, E_1=\frac{3}{4} \, ,\,\, E_2=-\frac{21}{16} \, ,\,\,
E_3=\frac{333}{64} \, ,\,\, E_4=-\frac{30885}{1024} \,\, .
\end{equation}
This coincides with the results obtained by using the standard
Rayleigh-Schr\"odinger perturbation method, with the advantage that no
information about the eigenstates of energy levels different from the one
being considered is required in order to obtain the terms of higher order.
The same can be done for other values of $b$.

Similar expansions can be found around any point other than $\lambda=0$. Let's
call $\lambda_{\alpha} = \lambda - \alpha$, then we can write
\begin{equation}
\left( -\frac{d^2}{dx^2} + x^a + \alpha x^b + \la x^b \right) \psi =
E \psi \,\, ,
\end{equation}
and now we can expand around $\lambda_{\alpha}=0$ (i.e. around
$\lambda=\alpha$). 
\begin{eqnarray}
E &=& \Ea_0+\Ea_1\la+\Ea_2\la^2 + \cdots \,\, ,  \\
\psi &=& \psia_0 +\psia_1\la +\psia_2\la^2 + \cdots  \,\, .
\end{eqnarray}

The following set of equations is obtained
\begin{eqnarray}
\La\psia_0 &=& \Ea_0 \psia_0 \label{PSa_DE1} \\
\La\psia_1 + x^b \psia_0 &=& \Ea_0 \psia_1 + \Ea_1 \psia_0 \,\, , \\
\La\psia_2 + x^b \psia_1 &=& \Ea_0 \psia_2 + \Ea_1 \psia_1 + \Ea_2 \psia_0 \,\, , \\
\vdots \phantom{x^b \psi_1}  && \phantom{E_0 \psi_1} \vdots \nonumber \\ 
\La\psia_n + x^b \psia_{n-1} &=& \sum_{k=0}^n \Ea_{n-k} \psia_k \,\, ,
\end{eqnarray}
where
\begin{equation}
\La = -\frac{d^2}{dx^2} + x^a + \alpha x^b \,\, .
\end{equation}
Clearly, equation (\ref{PSa_DE1}) will not have exact solutions for any values
of $a$ and $b$, so one is forced to find the coefficients
numerically. Equations (\ref{intE1}) and (\ref{intEn}) will still be valid,
but with the changes $\psi_k \rightarrow \psia_k$ and $E_k \rightarrow \Ea_k$,
i.e.,
\begin{equation}
\Ea_1 = \frac{\int_{-\infty}^{\infty} dx x^b (\psia_0)^2}{
      \int_{-\infty}^{\infty} dx (\psia_0)^2} \,\, 
\end{equation}
and
\begin{equation}
\Ea_n = \frac{\int_{-\infty}^{\infty} dx \left( x^b \psia_{n-1} 
                           -\sum_{k=1}^{n-1} \Ea_{n-k} \psia_k \right) \psia_0}{
      \int_{-\infty}^{\infty} dx \psi_0^2} \,\, .
\end{equation}

The coefficient $\Ea_k$ is actually the value of the $k$-th derivative of the
function $E(\lambda)$ evaluated at $\lambda = \alpha$. One might find these
derivatives directly, evaluating $E(\lambda)$ nearby $\lambda = \alpha$, for
example,
\begin{equation}
\Ea_1 = \frac{E(\lambda=\alpha)-E(\lambda=\alpha+\epsilon)}{\epsilon} \,\, .
\end{equation}
However, this way of finding the coefficients becomes relatively difficult for
higher derivatives, since then one needs to evaluate the function $E(\lambda)$
with increasing accuracy. The method proposed here can be viewed as an alternative, 
more accurate, easier and more efficient way to find these derivatives.

\section{Asymptotic Expansion}
\label{AsymptoticExpansion}

Doing the change of variables $x=\lambda^{-\frac{1}{2+b}}y$, 
and defining $\tl=\lambda^{-\frac{2+a}{2+b}}$ and 
$\tE=\lambda^{-\frac{2}{2+b}}E$, the Schr\"odinger equation becomes
\begin{equation}
\left(-\frac{d^2}{dy^2} + \tl y^a + y^b \right) \psi = \tE \psi \,\, .
\label{AsySE}
\end{equation}
One can expand now $\tE$ and $\psi$ in a similar way as before,
\begin{eqnarray}
\tE &=& \tE_0+\tE_1\tl+\tE_2\tl^2 + \cdots \,\, , \label{tE} \\ 
\psi &=& \tpsi_0 +\tpsi_1\tl +\tpsi_2\tl^2 + \cdots \,\, , 
\end{eqnarray}
Introducing this in equation (\ref{AsySE}) leads also to a system of
differential equations
\begin{eqnarray}
\tL\tpsi_0 &=& \tE_0 \tpsi_0 \,\, , \label{AsyDE1} \\
\tL\tpsi_1 + y^a \tpsi_0 &=& \tE_0 \tpsi_1 + \tE_1 \tpsi_0 \,\, , \label{AsyDE2} \\
\tL\tpsi_2 + y^a \tpsi_1 &=& \tE_0 \tpsi_2 + \tE_1 \tpsi_1 + \tE_2 \tpsi_0 \,\, , \\
\vdots \phantom{y^a \psi_1}  && \phantom{E_0 \psi_1} \vdots \nonumber \\
\tL\tpsi_n + y^a \tpsi_{n-1} &=& \sum_{k=0}^n \tE_{n-k} \tpsi_k \,\, , \label{AsyDEn}
\end{eqnarray}
where
\begin{equation}
\tL = -\frac{d^2}{dy^2} + y^b
\end{equation}

Rewriting the asymptotic expansion in terms of $\lambda$ instead of $\tl$, it
is clear that the form of the expansion depends on the particular potential to
be considered. In the case of the quartic anharmonic oscillator ($a=2$ and $b=4$), equation
(\ref{tE}) leads to
\begin{eqnarray}
E &=& \lambda^{1/3} \left(\tE_0+\frac{\tE_1}{\lambda^{2/3}}+
\frac{\tE_2}{\lambda^{4/3}}+\frac{\tE_3}{\lambda^2} + \cdots \right) 
\nonumber \\ 
&=& \lambda^{1/3} \sum_{k=0}^{\infty} \frac{\tE_{3k}}{\lambda^{2k}}
   +\lambda^{-1/3} \sum_{k=0}^{\infty} \frac{\tE_{3k+1}}{\lambda^{2k}} 
   +\frac{1}{\lambda} \sum_{k=0}^{\infty} \frac{\tE_{3k+2}}{\lambda^{2k}}
\end{eqnarray}
while in the case of the sextic anharmonic oscillator ($a=2$ and $b=6$), we obtain
\begin{eqnarray}
E &=& \lambda^{1/4} \left( \tE_0 + \frac{\tE_1}{\lambda^{1/2}} + \frac{\tE_2}{\lambda}+
\frac{\tE_3}{\lambda^{3/2}} + \frac{\tE_4}{\lambda^2} + \cdots \right) 
\nonumber \\
&=& \lambda^{1/4} \sum_{k=0}^{\infty} \frac{\tE_{2k}}{\lambda^{k}}
+ \lambda^{-1/4} \sum_{k=0}^{\infty} \frac{\tE_{2k+1}}{\lambda^{k}} \,\, .
\end{eqnarray}
In general, the expansions will have this structure, i.e.,
they can be divided in a few pieces, each one consisting in a series of
negative integer powers of $\lambda$, multiplied by a rational power of $\lambda$. For
this reason, the approximants that we will build must also be divided in a
similar way, in order to match the behavior of each piece. This will be seen explicitly
in the next two sections.

\section{Approximants for the quartic anharmonic oscillator}
\label{quartic}

For the quartic anharmonic oscillator \cite{R7,R8,A0,A1,A2}, the approximants for the energy
eigenvalues can be written in the following form
\begin{equation}
E_{\rm app}(\lambda) = (1+\mu \lambda)^{1/3}\frac{P_a(\lambda)}{Q(\lambda)}
                      +(1+\mu \lambda)^{-1/3}\frac{P_b(\lambda)}{Q(\lambda)}
                      +\frac{1}{1+\mu \lambda}\frac{P_c(\lambda)}{Q(\lambda)} \,\, ,
\label{Eapp}
\end{equation}
where
\begin{equation}
P_a(\lambda) = \sum_{k=0}^N a_k \lambda^k \,\, , \quad
P_b(\lambda) = \sum_{k=0}^N b_k \lambda^k \,\, , \quad
P_c(\lambda) = \sum_{k=0}^N c_k \lambda^k \,\, ,
\label{Ps}
\end{equation}
and
\begin{equation}
Q(\lambda) = 1 + \sum_{k=1}^N q_k \lambda^k \,\, ,
\label{Q}
\end{equation}
that is, the approximant is constructed using rational functions multiplied by
auxiliary ones, conveniently chosen in order to match the asymptotic behavior 
of the eigenvalues. Furthermore, since the power series is also going to be
used, it should be possible to Taylor-expand 
these functions around positive values of $\lambda$. It is for this last reason that the
auxiliary functions are not chosen directly as the factors of $\lambda^{1/3}$,
$\lambda^{-1/3}$ and $1/\lambda$ that appear multiplying each one of the three
pieces that make up the asymptotic expansion. Instead, we do the
change $\lambda \rightarrow 1+ \mu \lambda$ inside these roots, which, of course,
still gives the right behavior for $\lambda \rightarrow \infty$. An arbitrary
factor of $\mu$ has been included, which can be adjusted in order to improve
the precision of the approximant. 

With this choice of auxiliary functions the
degrees of the polynomials in the numerator must be the same as the ones in the
denominator. In principle, this can be done independently for each one of the three pieces
in equation (\ref{Eapp}), i.e. $P_a(\lambda)$, $P_b(\lambda)$ and
$P_c(\lambda)$ could be chosen with different degrees, and in that case,
different denominators matching the degree of each one of these polynomials
would be needed. For simplicity, a denominator $Q(\lambda)$ common to all
three pieces has been chosen, and so all polynomials have the same
degree. As it will be understood later, any other choice would lead to
a system of non-linear equations in the $p_k$'s and $q_k$'s, 
making the determination of the approximant unnecessarily complicated.

The coefficients of the polynomials in the approximant are found using the
power series, asymptotic expansion and the expansions around intermediate
points ($0< \alpha <\infty$), whose calculation was explained in the previous
two sections. One is free to choose as many
terms from each expansion as one desires, as long as the total number of terms
from all expansions equals the total number of coefficients in the
approximant. If the degree of the polynomials is $N$, the total number of
coefficients will be $4N+3$. In general, the approximants will have higher
precision for higher $N$. 

The values of the first few terms in the power series
(around $\lambda=0$) for the first three energy levels (labeled by $n$) 
of the quartic anharmonic
oscillator are shown in table \ref{Coeffx2x4Pow}, while the values of the first few terms in
the asymptotic expansion are shown in table \ref{Coeffx2x4Asy}. Notice that in accordance
with what was discussed in section \ref{PowerSeries}, the values of the coefficients for the
power series are exact. The values of the coefficients for the asymptotic
expansion were obtained by solving equations (\ref{AsyDE1})-(\ref{AsyDEn})
using the shooting method. For the ground state ($n=0$) and the second excited
level ($n=2$), the eigenfunctions are even in $x$, and as mentioned
before, so must be the functions $\tpsi_k$ (and the same applies to
$\psi_k$ and $\psia_k$), so the initial conditions used in those cases were $\tpsi_k(0)=1$
and $\tpsi'_k(0)=0$. For the first excited level the eigenfunction is odd in
$x$, so the conditions were $\tpsi_k(0)=0$ and $\tpsi'_k(0)=1$. One might feel
uneasy about the propagation of errors from one differential equation to the
next, but it can be checked numerically that the accuracy of the energy eigenvalues
for large values of $\lambda$ (or small values of $\tl$) improves as one
includes higher terms in the expansion, which gives us confidence that the
precision of the coefficients is acceptable.

\begin{table}
\begin{tabular}{|c|c|c|c|}
\hline
Coefficients &      $n=0$      &      $n=1$      &       $n=2$     \\ 
\hline \hline
    $E_0$   & 1            & 3              & 5              \\
    $E_1$   & 3/4          & 15/4           & 39/4           \\
    $E_2$   & -21/16       & -165/16        & -615/16        \\
    $E_3$   & 333/64       & 3915/64        & 20079/64       \\
    $E_4$   & -30885/1024  & -520485/1024   & -3576255/1024  \\
    $E_5$   & 916731/4096  & 21304485/4096  & 191998593/4096 \\
\hline
\end{tabular}
\caption{Exact coefficients of the power series for the first three energy
  levels of the quartic anharmonic oscillator.}
\label{Coeffx2x4Pow}
\end{table}

\begin{table}
\begin{tabular}{|c|c|c|c|}
\hline
Coefficients &      $n=0$      &      $n=1$      &       $n=2$     \\ 
\hline \hline
    $\tE_0$   & 1.060361944892  & 3.7996728480480 & 7.455697915983 \\
    $\tE_1$   & 0.362022935     & 0.901605953     & 1.244714261    \\
    $\tE_2$   & -0.034510565    & -0.057483095    & -0.046601602   \\
    $\tE_3$   & 0.005195593     & 0.005492673     & 0.000958945    \\
    $\tE_4$   & -0.000831127    & -0.000513914    & -0.000831127   \\
\hline
\end{tabular}
\caption{Coefficients of the asymptotic expansion for the eigenvalues of the
  quartic anharmonic oscillator obtained solving the
  differential equations using the shooting method.}
\label{Coeffx2x4Asy}
\end{table}

Let's choose a few intermediate points $\alpha_i$ ($i=1,2, \dots$), and let's take $n_i$ terms
from the expansion around each one of these points. Let's also take $n_0$ terms from the
power series (around $\lambda=0$) and $n_a$ terms from the asymptotic
expansion. It will be assumed that $\sum_i n_i+n_0+n_a=4N+3$. Using the power
series at $\lambda=0$ one can write
\begin{equation}
Q(\lambda)\sum^{n_0}_{k=0}E_k\lambda^k =
(1+\mu \lambda)^{1/3} P_a(\lambda) +
(1+\mu \lambda)^{-1/3} P_b(\lambda)+
\frac{1}{1+\mu \lambda} P_c(\lambda) \,\, ,
\end{equation}
Taylor-expanding each side of this equation in $\lambda$, and demanding it to be satisfied at every order up to
$\lambda^{n_0}$, one obtains a set of $n_0$ linear equations in the
coefficients of the approximant. Likewise, one can use the expansions at the
intermediate points, and doing the change $\lambda=\lambda_{\alpha_i}+\alpha_i$,
one can write
\begin{eqnarray}
\left(1 + \sum_{k=1}^N q_k (\lambda_{\alpha_i}+\alpha_i)^k \right)
\sum^{n_i}_{k=0}E^{\alpha_i}_k\lambda_{\alpha_i}^k &=&
(1+\mu (\lambda_{\alpha_i}+\alpha_i))^{1/3} \sum_{k=0}^N a_k
(\lambda_{\alpha_i}+\alpha_i)^k \nonumber \\
&& +
(1+\mu (\lambda_{\alpha_i}+\alpha_i))^{-1/3} \sum_{k=0}^N b_k
(\lambda_{\alpha_i}+\alpha_i)^k \nonumber \\
&& +
\frac{1}{1+\mu (\lambda_{\alpha_i}+\alpha_i)} \sum_{k=0}^N c_k
(\lambda_{\alpha_i}+\alpha_i)^k 
\end{eqnarray}
If one demands this equation to be satisfied at every order in
$\lambda_{\alpha_i}$ up to $\lambda_{\alpha_i}^{n_i}$, 
one obtains a set of $n_i$ linear equations in the
coefficients. Finally, one can use the asymptotic expansion. For this we need
to do the change $\lambda'=1/\lambda$, and match the expansion with the
approximant for each one of the three pieces in which it is divided. For
example, since 
\begin{equation}
(1+\mu \lambda)^{1/3} \frac{P_a(\lambda)}{Q(\lambda)} =
\lambda^{1/3} (\mu + \lambda')^{1/3} 
\frac{\sum_{k=0}^N a_k \lambda'^{N-k}}{1 + \sum_{k=1}^N q_k \lambda'^{N-k}}
\,\, ,
\end{equation}
one can compare the term multiplying $\lambda^{1/3}$ in the right hand side of
this equation with the term multiplying the same factor in the asymptotic
expansion. Doing the same also for the other two pieces leads to
\begin{eqnarray}
\left( 1 + \sum_{k=1}^N q_k \lambda'^{N-k} \right) 
\sum_{k=0}^{\infty} \tE_{3k}\lambda'^{2k} &=&
(\lambda'+\mu)^{1/3} \sum_{k=0}^N a_k \lambda'^{N-k} \,\, ,\\
\left( 1 + \sum_{k=1}^N q_k \lambda'^{N-k} \right) 
\sum_{k=0}^{\infty} \tE_{3k+1}\lambda'^{2k} &=&
(\lambda'+\mu)^{-1/3} \sum_{k=0}^N b_k \lambda'^{N-k} \,\, , \\
\left( 1 + \sum_{k=1}^N q_k \lambda'^{N-k} \right) 
\sum_{k=0}^{\infty} \tE_{3k+2}\lambda'^{2k} &=&
\frac{1}{\lambda'+\mu}\sum_{k=0}^N c_k \lambda'^{N-k} \,\, .
\end{eqnarray}
Here the number of terms taken in each expansion is determined by
$n_a$, that is, one would not allow any $\tE_k$ with $k>n_a$ in
the sums.
In this way, one gets a set of $n_a$ linear equations for the coefficients
of the approximant.

In table \ref{x4d3}, the values of the coefficients of the approximants are
shown for the first three energy levels, using polynomials of degree three. There
are fifteen coefficients in each approximant, and they were obtained using the
first five terms of the power series (around $\lambda=0$), the first five
terms of the asymptotic expansion, and the first term of the series around
$\lambda=0.5$, $\lambda=1$, $\lambda=2$, $\lambda=5$ and $\lambda=20$ (which
are shown for the three energy levels in table \ref{Interx4d3}). This
means that we are only using the exact energy eigenvalue around these
intermediate points, and forcing the approximant built with the power series and
asymptotic expansion to furthermore coincide with these ``exact'' eigenvalues at
these points. This not only brings the relative error of the
approximant at these points down to zero (they become nodes of the relative
error as a function of $\lambda$), but also helps to decrease the
error in between these points. The relative error is defined using as target the
eigenvalues obtained numerically through the shooting method, i.e., the
relative error is given by
\begin{equation}
\frac{|E_{\rm app}- E_{\rm shooting}|}{E_{\rm shooting}} \,\, .
\end{equation} 

\begin{table}
\begin{tabular}{|c|c|c|c|}
\hline
        Coefficients      &      $n=0$      &      $n=1$       &       $n=2$     \\ 
\hline \hline
    $E_0(\lambda=1/2)$   & 1.241854043136  & 4.051932338617   & 7.396900686938  \\
    $E_0(\lambda=1)$     & 1.392351580103  & 4.64881282723    & 8.655049982254  \\
    $E_0(\lambda=2)$     & 1.607541348124  & 5.475784646286   & 10.358583364736 \\
    $E_0(\lambda=5)$     & 2.018340657447  & 7.013479298703   & 13.467730394819 \\
    $E_0(\lambda=20)$    & 3.009944947791  & 10.643215959124  & 20.694110927154 \\
\hline
\end{tabular}
\caption{First coefficient (energy eigenvalues) of the series at different
  intermediate points for the first three energy levels with $V(x)=x^2+\lambda
  x^4$. These values were obtained using the shooting method.}
\label{Interx4d3}
\end{table}

\begin{table}[h]
\begin{tabular}{|c|c|c|c|}
\hline
Coefficients &      $n=0$      &      $n=1$      &       $n=2$     \\ 
\hline \hline
    $a_0$   &  -235.587774594 &  3.26113271857  &  -46.3903727540 \\
    $a_1$   &  129.192528081  &  45.7861842084  &  118.622015136  \\
    $a_2$   &  819.219808968  &  347.592601172  &  906.778841942  \\
    $a_3$   &  4083.20247083  &  1023.55148495  &  3353.40199807  \\
    $b_0$   &  49.9309955808  &  4.80222464913  &  8.35806073416  \\
    $b_1$   &  374.551951382  &  43.2593054339  &  113.555378592  \\
    $b_2$   &  1181.63222463  &  259.439145729  &  536.540936096  \\
    $b_3$   &  2212.93937770  &  385.537761596  &  888.696857827  \\
    $c_0$   &  186.656779013  &  -5.06335736770 &  43.0323120198  \\
    $c_1$   &  208.866219507  &  -20.7666223608 &  48.6886778830  \\
    $c_2$   &  -423.418335743 &  -35.6233569984 &  -59.8483255497 \\
    $c_3$   &  -334.866518168 &  -39.0190739652 &  -52.8167275564 \\
    $q_1$   &  148.201294158  &  24.5427291322  &  29.7104983824  \\
    $q_2$   &  1782.00574019  &  171.823102857  &  247.681714807  \\
    $q_3$   &  4851.65727491  &  339.396077796  &  566.683604101  \\
\hline
\end{tabular}
\caption{Coefficients for the approximants of the first three energy
  eigenvalues of $V(x)=x^2+\lambda x^4$, using polynomials of degree 3.}
\label{x4d3}
\end{table}

The highest relative error with these approximants was obtained for small
values of $\lambda$. Specifically, the maximum relative error was obtained
around $\lambda \approx 0.2$. In the case of the ground state, the highest
relative error was
\begin{equation}
\left. \frac{|E_{\rm app}- E_{\rm shooting}|}{E_{\rm shooting}}
\right|_{\lambda=0.17} = 1.05 \times 10^{-6} \,\, .
\end{equation} 
The relative error decreases rapidly for smaller values of $\lambda$, and of
course, it also decreases when $\lambda$ increases until it finds the next
node at $\lambda=0.5$. After that, the relative error never becomes higher
than $2 \times 10^{-7}$. For the first and second excited level,
the maximum error around $\lambda \approx 0.2$, was about $8 \times 10^{-7}$
and $2.4 \times 10^{-6}$, respectively, and after the node at $\lambda=0.5$
this error is never higher than $4 \times 10^{-8}$. In fact, the relative
error decreases quite rapidly in the case of the first and second excited levels
for large values of $\lambda$, although it does so more slowly in the case of
the ground state.

In all these approximants, we chose $\mu=2$. This parameter is arbitrary
except for one restriction: The approximants should not have any defects, that
is, there should not be any poles in the approximant (positive roots of $Q(\lambda)$)
with the corresponding nearby zeros. Notice in table \ref{x4d3} that with this
choice of $\mu$ all of the coefficients of $Q(\lambda)$ are positive, which
will, of course, guarantee that it has no roots for $\lambda>0$. 
Other choices of $\mu$ may lead to
mixed negative and positive coefficients in $Q(\lambda)$, which will in
general lead to positive real roots in this polynomial. Other than that, there
is no other restriction in $\mu$. Among all of the values of $\mu$ that allow
to keep the approximant free of defects, one is free to choose the one that
minimizes the relative errors.

Other than improving the numerical method used to obtain the coefficients of the
expansions, there are several ways in which the maximum relative error of the 
approximants can be decreased for all energy levels. The
easiest one is to move one of the nodes. For example, one may choose the
approximant to have a node at $\lambda=0.2$ instead of $\lambda=0.5$. If this
is done, it can be seen that the maximum relative error is reduced by about a
half for all energy levels. Another possibility is to use the
derivatives of $E(\lambda)$ at some of the intermediate points. Finally, one
may try an approximant of higher degree, allowing it
to coincide with the values of $E(\lambda)$ and its
derivatives at more points. This last possibility is studied in the next example.

\section{Approximants for the sextic anharmonic oscillator}
\label{sextic}

For the sextic anharmonic oscillator \cite{R9,R10}, the asymptotic expansion consists
of only two pieces, so the approximant can be written as

\begin{equation}
E_{\rm app}(\lambda) = (1+\mu \lambda)^{1/4}\frac{P_a(\lambda)}{Q(\lambda)}
                      +(1+\mu \lambda)^{-1/4}\frac{P_b(\lambda)}{Q(\lambda)} \,\, ,
\end{equation}
where $P_a(\lambda)$, $P_b(\lambda)$ and $Q(\lambda)$ are given in equations
(\ref{Ps}) and (\ref{Q}). The corresponding coefficients $a_k$, $b_k$ and
$q_k$ are found in a similar way as we did for the quartic anharmonic
oscillator. For approximants of degree $N$, we will have $3N+2$
coefficients, so taking $n_0$ terms from the power series, $n_i$ terms from the
series at the $i$-th intermediate point, and $n_a$ terms from the
asymptotic expansion, we should have $\sum_i n_i+n_0+n_a=3N+2$,
and the coefficients $a_k$, $b_k$ and $q_k$ can be determined using 
equations derived by matching powers in $\lambda$ in the equation
\begin{equation}
Q(\lambda)\sum^{n_0}_{k=0}E_k\lambda^k =
(1+\mu \lambda)^{1/4} P_a(\lambda) +
(1+\mu \lambda)^{-1/4} P_c(\lambda) \,\, ,
\end{equation}
matching powers in $\lambda_{\alpha_i}$ using
\begin{eqnarray}
\left(1 + \sum_{k=1}^N q_k (\lambda_{\alpha_i}+\alpha_i)^k \right)
\sum^{n_i}_{k=0}E^{\alpha_i}_k\lambda_{\alpha_i}^k &=&
(1+\mu (\lambda_{\alpha_i}+\alpha_i))^{1/4} \sum_{k=0}^N a_k
(\lambda_{\alpha_i}+\alpha_i)^k \nonumber \\
&& +
(1+\mu (\lambda_{\alpha_i}+\alpha_i))^{-1/4} \sum_{k=0}^N b_k
(\lambda_{\alpha_i}+\alpha_i)^k \nonumber \,\, ,
\end{eqnarray}
and matching powers in $\lambda'$ in
\begin{eqnarray}
\left( 1 + \sum_{k=1}^N q_k \lambda'^{N-k} \right) 
\sum_{k=0}^{\infty} \tE_{2k}\lambda'^{k} &=&
(\lambda'+\mu)^{1/4} \sum_{k=0}^N a_k \lambda'^{N-k} \,\, ,\\
\left( 1 + \sum_{k=1}^N q_k \lambda'^{N-k} \right) 
\sum_{k=0}^{\infty} \tE_{2k+1}\lambda'^{k} &=&
(\lambda'+\mu)^{-1/4} \sum_{k=0}^N b_k \lambda'^{N-k} \,\, . \\
\end{eqnarray}

The first few coefficients of the power series at $\lambda=0$ and the asymptotic
expansion, obtained using the systems of
differential equations described in sections \ref{PowerSeries} and
\ref{AsymptoticExpansion} are shown in tables \ref{Coeffx2x6Pow} and
\ref{Coeffx2x6Asy}, respectively. As expected, the coefficients of the power
series are exact, and the coefficients of the asymptotic expansion are
obtained numerically.

\begin{table}
\begin{tabular}{|c|c|c|c|}
\hline
Coefficients &        $n=0$      &        $n=1$        &         $n=2$        \\ 
\hline \hline
    $E_0$   & 1                 & 3                   & 5                     \\
    $E_1$   & 15/8              & 105/8               & 375/8                 \\
    $E_2$   & -3495/128         & -47145/128          & -295095/128           \\
    $E_3$   & 1239675/1024      & 27817125/1024       & 276931275/1024        \\
    $E_4$   & -3342323355/32768 & -110913018405/32768 & -1626954534555/32768  \\
\hline
\end{tabular}
\caption{Exact coefficients of the power series for the first three energy
  levels of $V(x)=x^2 + \lambda x^6$.}
\label{Coeffx2x6Pow}
\end{table}

\begin{table}
\begin{tabular}{|c|c|c|c|}
\hline
Coefficients &      $n=0$        &       $n=1$      &       $n=2$     \\ 
\hline \hline
    $\tE_0$   & 1.144802430723  & 4.338598612643   & 9.073084583078 \\
    $\tE_1$   & 0.307920324     & 0.718220191      & 0.904435602    \\
    $\tE_2$   & -0.018541674    & -0.024395762     & -0.010249211   \\
    $\tE_3$   & 0.001559745     & 0.00099946565    & -0.000749318   \\
    $\tE_4$   & -0.000123969    & -0.000026329     & 0.000107946    \\
\hline
\end{tabular}
\caption{Coefficients of the asymptotic expansion obtained solving the
  differential equations using the shooting method for $V(x)=x^2 + \lambda x^6$.}
\label{Coeffx2x6Asy}
\end{table}

The degree of the polynomials in the approximant was first chosen to be
$N=5$. The coefficients of the approximants were calculated choosing different
intermediate points (nodes) for the first three energy levels, together with
the first four terms of the power series and the first five terms of the
asymptotic expansion. For the
ground state, the intermediate points were $\lambda=0.1$, $\lambda=0.2$,
$\lambda=0.5$, $\lambda=1$, $\lambda=2$, $\lambda=5$ and $\lambda=10$, and
only the energy eigenvalues at those point were used, i.e., we didn't pick any
of the derivatives at these points. Furthermore, we chose $\mu=1/2$. 
For the first and second excited level, we used $E^1_0$, $E^2_0$, $E^5_0$, 
$E^{10}_0$, $E^{20}_0$, $E^{0.1}_0$, $E^{0.1}_1$ 
and $E^{0.01}_0$, with $\mu=0.95$. Notice that in these cases, the derivative at
$\lambda=0.1$, i.e., $E^{0.1}_1$, has also been used. With these
choices, the values of the coefficients in the approximants are given in table
\ref{x6d5}. With these approximants, the highest relative errors were obtained
around small values of $\lambda$. Specifically, at $\lambda=0.014$ the
relative error for the ground state eigenvalue is $2.55 \times 10^{-5}$. For even
smaller values of $\lambda$, this error decreases rapidly. For $1<\lambda<5$
that maximum error is $1.3 \times 10^{-7}$, and for $\lambda>5$ the maximum
error was $7 \times 10^{-8}$, decreasing rapidly for large values of
$\lambda$.

\begin{table}[h]
\begin{tabular}{|c|c|c|c|}
\hline
Coefficients &        $n=0$      &      $n=1$      &       $n=2$     \\ 
\hline \hline
    $a_0$   &  -228343.425175410234  &  -1455854.05235538211  &  -1636184.16769173318 \\
    $a_1$   &  -51504.1866802877753  &  -523067.617460085793  &  -319382.170585718291 \\
    $a_2$   &  -3246.30836836582236  &  -82992.9062424631909  &  -76246.2680853710652 \\
    $a_3$   &  17641.2930769775453   &  564652.836765077742   &  1736863.85800628006  \\
    $a_4$   &  40455.7925641666977   &  2737870.84947363967   &  8425191.38724236837 \\
    $a_5$   &  25845.5374941939783   &  3765927.01323457233   &  10377310.4646700931 \\
    $b_0$   &  228344.425175410234   &  1455857.05235538211   &  1636189.16769173318 \\
    $b_1$   &  108718.883825186570   &  1215348.76068812833   &  1098149.56754718475 \\
    $b_2$   &  12471.9577982789445   &  211783.949326608282   &  156879.990271492461 \\
    $b_3$   &  8971.90069757654071   &  247389.200455627253   &  461008.702437356268 \\
    $b_4$   &  12714.5141045106691   &  765825.922719107915   &  1349548.74023090975 \\
    $b_5$   &  4915.62090727820199   &  607633.680665566723   &  1008252.48843727777 \\
    $q_1$   &  126.840851046236208   &  245.543619745344247   &  306.370961578640294 \\
    $q_2$   &  3258.74684448027568   &  13846.7357289482276   &  20215.7966313260916 \\
    $q_3$   &  21339.1202042371419   &  208223.456783358953   &  314306.763606766761 \\
    $q_4$   &  39515.8524539180998   &  853339.703125454811   &  1215186.64688146943 \\
    $q_5$   &  18984.4284445198520   &  856945.745673868394   &  1129173.69341497105 \\
\hline
\end{tabular}
\caption{Coefficients for the approximants of the first three energy
  eigenvalues of $V(x)=x^2+\lambda x^6$, using polynomials of degree 5.}
\label{x6d5}
\end{table}

As it can be seen, in both the quartic and sextic anharmonic oscillators,
the region of $\lambda$ where it is more difficult to
achieve high accuracy is for $\lambda<0.5$. We tried also
approximants of degree 6 for the sextic anharmonic oscillator, and it
was possible to reduce the relative error in this region by a factor of
1/2. The coefficients of the approximants can be found in table
\ref{x6d6}. For $n=0,2$, we used $\mu=1/2$, while for $n=1$, we used
$\mu=1$. For $n=0$ the approximant was built using the first four terms of the
power series around $\lambda=0$ and the first five terms of the asymptotic
series, together with $E^1_0$, $E^2_0$, $E^5_0$, $E^{10}_0$, $E^{20}_0$,
$E^{1/2}_0$,  $E^{1/2}_1$, $E^{0.2}_0$, $E^{0.1}_0$, $E^{0.1}_1$
and $E^{0.01}_0$. For $n=1,2$, the same terms were taken, except that the
fourth term of the power series around $\lambda=0$ was replaced by
$E^{1/2}_2$ (i.e. the second derivative at $\lambda=1/2$). The improvement in 
the accuracy of the approximants for $\lambda<0.5$ is related to having used
more terms in the expansions around points in this region, including first and
second derivatives.

The fact that
the approximants become less accurate for small values of $\lambda$ may be related to the
analytic properties of the exact function $E(\lambda)$ \cite{R7}, although
this is far from clear. As it is well known,
the perturbative expansion (i.e., the power series around $\lambda=0$) is divergent for any $\lambda \neq
0$. However, in previous works where quasi-rational approximants have been
used, it has been found that the main factor determining the accuracy of
an approximant is the accuracy of the coefficients in the expansions used to
build it. For
example, in references \cite{MC1} and \cite{MC2} series with radius of convergence equal to zero
were used, yet the approximants were very accurate because the coefficients of
these series were determined with very high precision.

\begin{table}[h]
\begin{tabular}{|c|c|c|c|}
\hline
Coefficients &      $n=0$      &      $n=1$      &       $n=2$     \\ 
\hline \hline
    $a_0$   &  10662630.6230456957   &  -4140725.82498950878  &  -190986117.577446118 \\
    $a_1$   &  -792418.407061574860  &  -3392187.29848733455  &  -96347617.9985940628 \\
    $a_2$   &  -770199.048057820101  &  -691764.158783859149  &  -11188867.9611306088 \\
    $a_3$   &  -30901.7803416591912  &   352514.275425898392  &   582001.151582893834 \\
    $a_4$   &  663802.804751486527   &   1809854.68977494829  &   4102541.85172044638 \\
    $a_5$   &  1304251.65718790436   &   3042011.16842997607  &   6466260.98769018109 \\
    $a_6$   &  768824.185718239332   &   443056.032781395005  &   2682339.11910005685 \\
    $b_0$   &  -10662629.6230456957  &   4140728.82498950878  &   190986122.577446118 \\
    $b_1$   &  -18730301.9075998957  &   5463158.61490945598  &   144095348.884473491 \\
    $b_2$   &  1312291.68171628629   &   1901041.02272365822  &   29373600.9144712753 \\
    $b_3$   &  289403.682268640790   &   314656.648970897322  &   1721958.42036808095 \\
    $b_4$   &  312257.358605665963   &   547382.135140547781  &   658971.157529230082 \\
    $b_5$   &  397391.953291475945   &   540767.205553174330  &   644912.329241756943 \\
    $b_6$   &  146224.401105478984   &   73344.3715121889572  &   189069.456358376232 \\
    $q_1$   &  207.057939859483339   &   198.176309122347673  &   230.798303579776449 \\
    $q_2$   &  10393.8027931755210   &   9463.18894774927319  &   11496.2792134605234 \\
    $q_3$   &  157632.217425587120   &   133763.388618245199  &   150748.225437877810 \\
    $q_4$   &  775120.602962958890   &   586955.795810181245  &   587462.218227788458 \\
    $q_5$   &  1249527.64792219401   &   727254.753026886137  &   723876.015243016534 \\
    $q_6$   &  564727.576025957156   &   102119.617954584846  &   248600.057576103110 \\
\hline
\end{tabular}
\caption{Coefficients for the approximants of the first three energy
  eigenvalues of $V(x)=x^2+\lambda x^6$, using polynomials of degree 6.}
\label{x6d6}
\end{table}

\section{Conclusions}

In this paper, it has been shown that accurate approximants for the energy
eigenvalues of potentials of the form $V(x)=Ax^a+Bx^b$ can be found using a
multi-point quasi-rational approximation technique. The approximants are
constructed using rational functions, together with auxiliary functions
introduced to be able to reproduce the behavior of the eigenvalues for large
$\lambda$. The coefficients of the rational functions are found using the
power series of the eigenvalues, not only at $\lambda=0$, but also for
arbitrary finite values of $\lambda$, together with the asymptotic
expansion. These expansions are found using a system of differential
equations, which in the case of the power series at $\lambda=0$, represents an
alternative way to find the perturbative expansion. As
examples, approximants for the lowest energy levels of the quartic and sextic
anharmonic oscillators were obtained. The approximants were fairly simple,
since the degree of the polynomials used was not too high. In particular, for
the quartic anharmonic oscillator, it was shown that it is possible to obtain
approximants with polynomials of degree 3 for which the relative error is not
higher than $\sim 3 \times 10^{-6}$. In the case of the sextic anharmonic,
polynomials of degree 5 and 6 were tried, and it was shown that in the second
case it was possible to improve the accuracy for $\lambda<0.5$ while
maintaining the accuracy for larger values of $\lambda$

In this technique, one has a lot of freedom choosing
the intermediate points, as well as how many terms from each one of the series to
take. This gives a lot of possibilities to try to reduce relative errors, and
it would be interesting to study if it is possible to do this systematically.
It would also be interesting to try to find methods that allow to improve the accuracy
of the coefficients in the expansions. For example, if better numerical solutions 
for the expansion functions can be found, this should lead to better coefficients 
and therefore to better approximants. As it was mentioned before, experience with quasi-rational
approximants has shown that the accuracy of the 
coefficients in the expansions is the main factor influencing the accuracy of multi-point 
quasi-rational approximations. For simplicity, here we have limited ourselves
to the use of the shooting method to solve the systems of differential
equations. On the other hand, the first equation in each one of the systems, 
i.e., equations (\ref{PS_DE1}), (\ref{PSa_DE1}) and
(\ref{AsyDE1}) are just regular Schr\"odinger equations, and many methods have
been developed that allow to obtain the corresponding eigenvalues with very high precision 
\cite{R4,A0,A1,A2,A3,A4,A5,A6,A7,A8,A9}. 
In fact, we could have used any of these methods to
obtain the coefficients $E_0^{\alpha}$ and $\tE_0$, but it is not clear how to
extend these methods to solve the remaning equations in the systems.
In the future, we plan to study these issues in more detail, 
and apply this technique to other potentials of interest,
both in physics as well as chemistry.

\end{document}